\documentclass[]{article}  
\usepackage{amsmath,amsfonts,amssymb}
\usepackage{graphicx}
\usepackage{fancyhdr}
\newcommand{\be}{\begin{equation}}
\newcommand{\ee}{\end{equation}}
\newcommand{\lo}{{\cal L}_0}

\title{Turbulence monitoring at the Plateau de Calern with the GDIMM instrument} 
\date{Eric Aristidi\footnote{UMR 7293, Lagrange, Universit\'e de Nice-Sophia Antipolis, CNRS, OCA, Parc Valrose F-06108 Nice Cedex 2, France}, Yan Fantei-Caujolle$^1$, Julien Chab\'e\footnote{Universit\'e C\^ote d'Azur, Observatoire de la C\^ote d'Azur, CNRS, IRD, G\'eoazur, 2130 route de
l'Observatoire, 06460 Caussols, France}, Catherine Renaud$^1$, Aziz Ziad$^1$, Malak Ben Rahhal$^1$}

\begin{document} 
  
  \maketitle 

\begin{abstract}
We present some statistics of turbulence monitoring at the Plateau de Calern (France), with the Generalised Differential Image Motion Monitor (GDIMM). This instrument allows to measure integrated parameters of the atmospheric turbulence, i.e. seeing, isoplanatic angle, coherence time and outer scale, with 2 minutes time resolution. It is running routinely since November 2015 and is now fully automatic. A large dataset has been collected, leading to the first statistics of turbulence above the Plateau de Calern.
\end{abstract}


\section{INTRODUCTION}

\label{par:intro}  
The Generalized Differential Image Monitor (GDIMM) is an instrument dedicated to the monitoring of atmospheric turbulence integrated parameters. It allows  measurements of the seeing $\epsilon$, the isoplanatic angle $\theta_0$, the coherence time $\tau_0$ and the spatial coherence outer scale $\lo$. This instrument was developped in the early 2010s to replace the aging Generalized Seeing Monitor (GSM) which had become a standart in the field of site testing (see [\cite{Ziad00}] and references therein). The GDIMM was first proposed in 2014 in a former SPIE meeting\cite{Aristidi14}. Based on a small commercial telescope (diameter 11in.), it is a compact and easy to use instrument, using original techniques to derive turbulence parameters. It is controlled by a dedicated software and has now attained a high level of automatisation. The present paper is an update of previous presentations\cite{Aristidi14, Chabe16, Ziad17}. It aims at giving the improvements and the present status of the instrument, and some statistics of turbulence above the Plateau de Calern.

\begin{figure}
\begin{center}
\includegraphics[width=9cm]{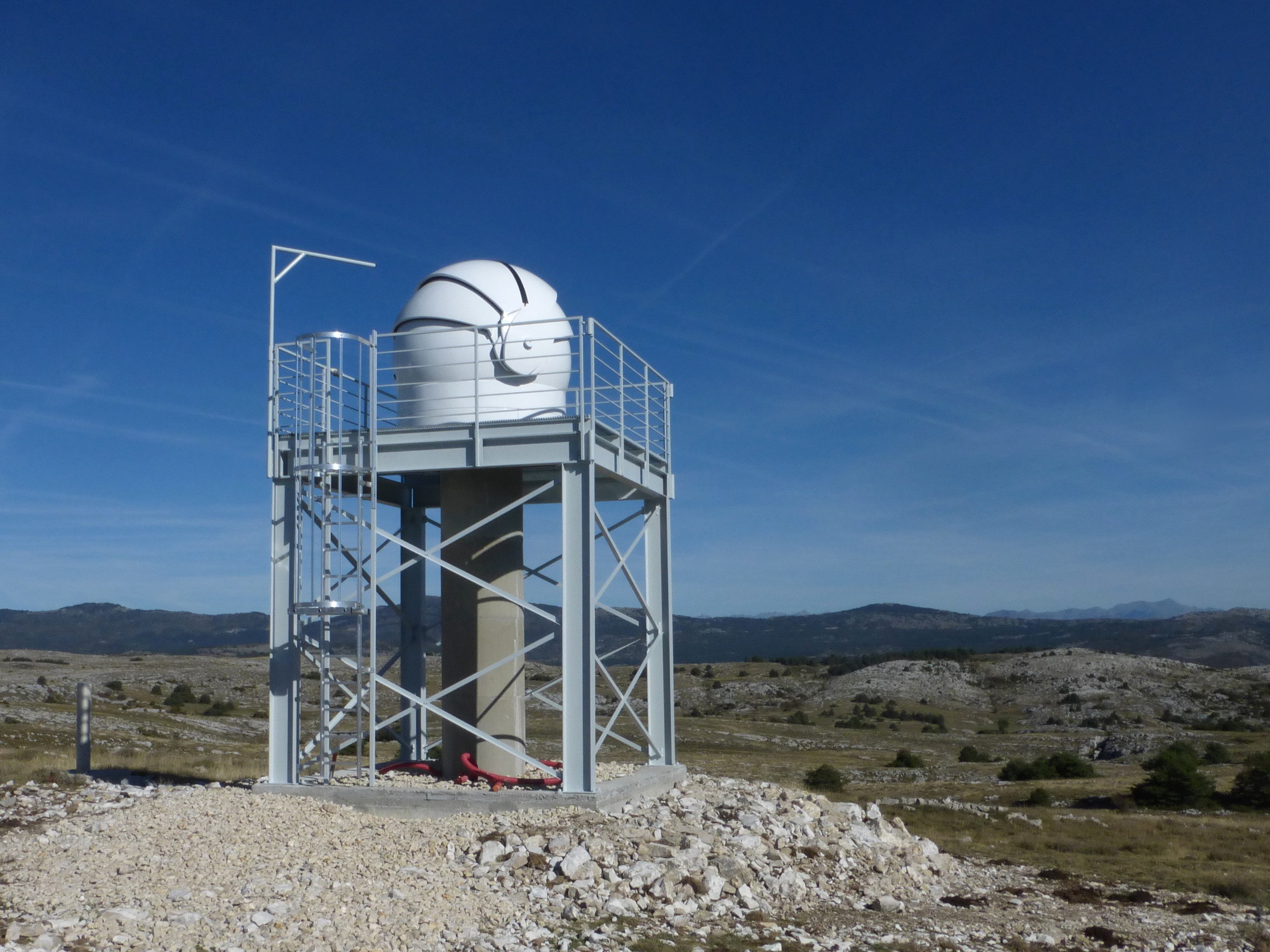}
\end{center}
\caption{The GDIMM tower just after its installation on the Plateau de Calern in September 2015}
\label{fig:photogdimm}
\end{figure}

\begin{figure}
\begin{center}
\includegraphics[width=12cm]{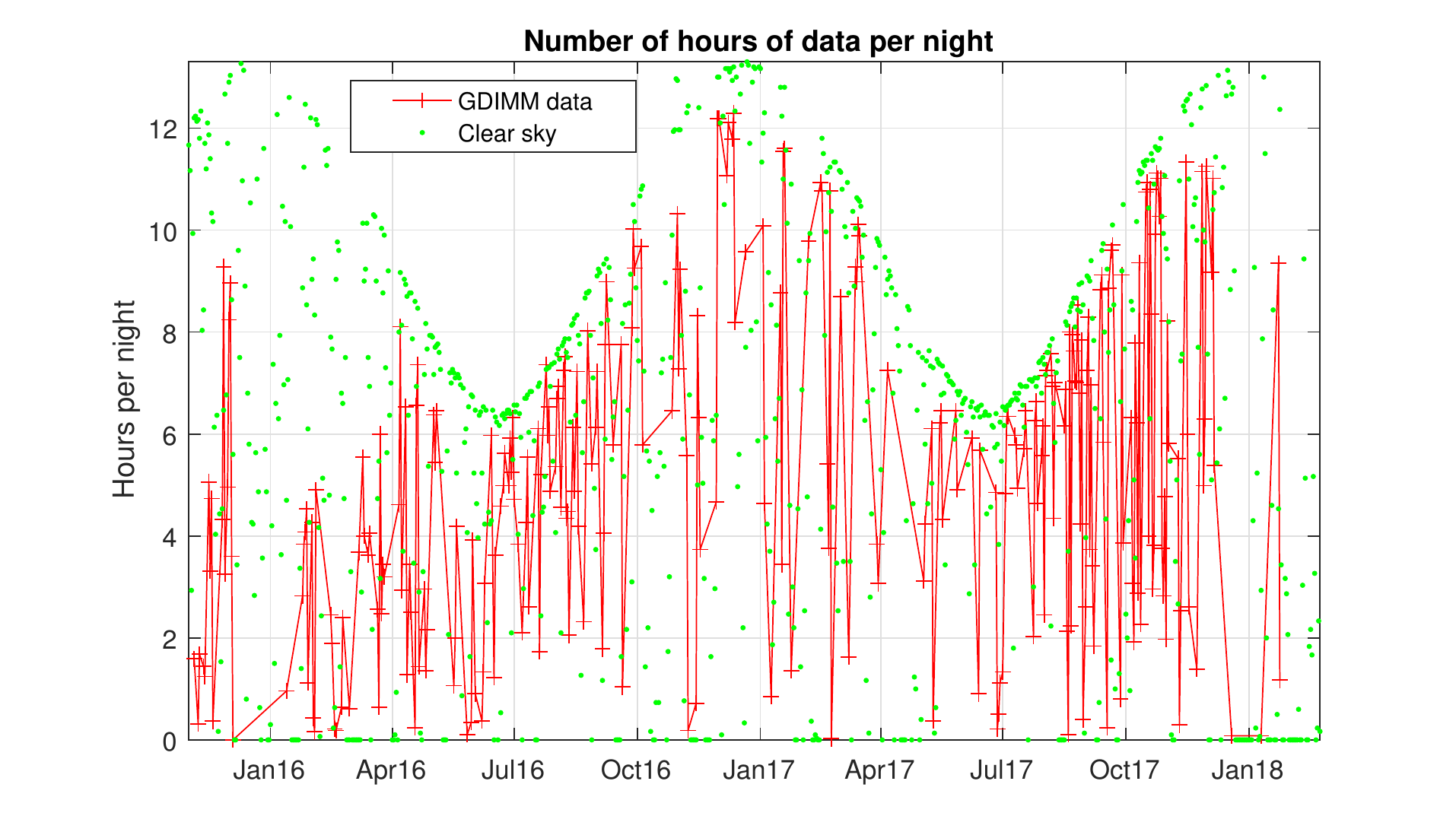}
\end{center}
\caption{Number of hours of GDIMM data per night since the beginning of GDIMM exploitation. Green dots are the number of hours of clear sky measured by the All-Sky camera.
}\label{fig:statgdimm}
\end{figure}

GDIMM is indeed a part of the Calern Atmospheric Turbulence Station (CATS)\cite{Chabe16, Ziad18}, who was developped as a site monitoring facility at the Plateau de Calern (France), in connexion with the laser telemetry MeO station\cite{Samain08}.  The other instruments composing CATS are the Profiler of Moon Limb\cite{Benrahhal18, Catala16, Ziad13}, an All-Sky camera providing the cloud coverage and a meteo station. More details on CATS are given in [\cite{Ziad18}].
GDIMM first data were collected in 2013 and 2014 at the Plateau de Calern, and at the Mont-Gros observatory (Nice, France), and were the base for our first paper\cite{Aristidi14}. Simultaneous observations with GSM were made during a few nights in the summer of 2014 and 2015, to cross-check the data and affinate the processing. The instrument was moved in November 2015 on the top of a 4m~high concrete pillar, and protected by a dome. The dome is attached to a metallic platform and disconnected from the GDIMM pillar to avoid vibrations (see Fig.~\ref{fig:photogdimm}). From November to now on, a lot was done to make this instrument fully automatic, using informations from the meteo station and the All-Sky camera to decide whether or not it is possible to observe. As the automatisation progressed, the number of daily collected data became larger. This is illustrated by Fig~\ref{fig:statgdimm}, which represents the number of hours of GDIMM observations as a function of time since November 2015. The green dots visible on the graph are the number of hours of clear sky, measured simultaneously by the All-Sky camera (the green dots are bounded by a sinusoid representing the duration of the night). There was a clear improvement during the first half of the year 2016, resulting in a reduction time loss due to instrumental problems. The instrument is now running at its full capacity, and under stable meteo conditions the efficiency can attain almost 100\% (the efficiency is defined as the ratio between the number of hours of data and the duration of the night). Real-time measurements are accessible worldwide from a dedicated website {\tt cats.oca.eu}.

\section{OBSERVATIONS AND DATA PROCESSING}
\subsection{Description of the instrument}
GDIMM is an evolution of the popular DIMM proposed in the early 90s\cite{Sarazinroddier90} to estimate the seeing from differential motion of two sub-images produced by a telescope using a 2~apertures pupil mask. GDIMM pupil mask has 3 apertures (numbered as 1, 2 and 3) of diameter $D_1=D_2=6$cm and $D_3=10$cm (with a central obstruction of 4cm). Sub-pupils 1 and 2 are separated by $B=20$cm. The telescope itself is a Celestron C11 placed onto an equatorial mount Astro-Physics AP900. A wide field finder (4 degrees) equiped with a webcam allows to point and center the target stars on the science camera. The limiting magnitude is about 2 with an exposure time of 5~ms. A detailed description is given in previous papers\cite{Aristidi14, Chabe16}.


\subsection{GDIMM daily acquisition sequence}
\label{par:obs}
GDIMM was designed to operate in a fully automatic mode. Routine observations are performed every night, according to the following procedure. 
\begin{itemize}
\item Start operations when the Sun elevation is below -3$^\circ$ 
\item Check for meteo station. Authorization is given to open the dome if the wind speed and the humidity are lower than a threshold (current values are 8.5 m/s for the wind speed, and 85\% for humidity), and if the difference between the temperature and the dewpoint is greater than 0.5$^\circ$C.
\item Wait for the night (Sun elevation below -6$^\circ$)
\item Check the All-Sky camera, and start acquisition sequence if the clear sky fraction is greater than 60\%
\item Select a star from an internal list of bright stars (magnitude$<$2). The star selection process is based upon an algorithm taking into account several parameters (star magnitude, duration of observability (i.e. zenith distance $<50^\circ$) to give a score to each star. The star having the best ranking is selected.
\item Point the star and center it on the finder camera. If no star is detected, go to the next star in the ranking list.
\item Search for star image on the science camera. If no star is detected, make a spiral search around the pointed position in the finder camera
\item Start acquisitions images cubes,  online processing is performed in real time (more details are given in [\cite{Aristidi14}]) and the turbulence parameters are sent to a web server available for users on the observatory.
\item Change star when it moves out the observability zone, or in case of signal loss (possible clouds).
\item During the whole sequence, meteo conditions and sky coverage are periodially checked. Observations are stopped if conditions degradate. A supplementary hardware security is given by the inverter protecting the acquisition PC: instrument is automatically closed in case of power failure.
\item Sequence is stopped at the end of the night, and dome is closed when the Sun elevation is above -3$^\circ$, awaiting the following night.
\end{itemize}

\subsection{Advances in the data processing}
\label{par:dataproc}

\subsubsection{Seeing and isoplanatic angle}
Calculation of the seeing and the isoplanatic angle was presented in details in [\cite{Aristidi14}]. Seeing estimations are based on variances of the photocenter difference of images produced by sub-pupils 1 and 2 (two values of this differential variance are computed for directions parallel and perpendicular to the pupil separation). Isoplanatic angle is derived from scintillation measurements on images produced by the pupil 3, using the model from Loos and Hogge\cite{Looshogge79}. To compensate for the effect of finite exposure time\cite{Tokovinin02, Aristidi14}, image cubes are composed of 2 consecutive sequences of $N=1024$ images taken with exposure times of $T=5$ms and $2T$. To gain speed, only a small region of the detector around the star images is actually read. The current frame rate is about 100 frames/second. 
Simultaneous measurements of the seeing and the isoplanatic angle make it possible to derive the average turbulence altitude defined by Roddier \cite{Roddier82} as
\be
\bar{h}=0.31 \frac{r_0}{\theta_0} 
\ee
with $r_0=0.98 \frac{\lambda}{\epsilon}$ the Fried parameter. Some statistics for $\bar h$ at Calern are presented in Section~\ref{par:results}. 
\subsubsection{Outer scale}
In [\cite{Aristidi14}] we proposed to make use of variances of the absolute motions of sub-images to estimate the outer scale $\lo$. These absolute variances (in square radians) are given by\cite{Conan00}
\be
\label{eq:varabs}
\sigma_D^2=0.17\, \lambda^2 r_0^{-5/3}\, (D^{-1/3}-1.525 \lo^{-1/3})
\ee
Since we have two different pupil diameters ($D_1=6$cm and $D_3=10$cm), it is possible to derive the outer scale from the following ratio:
\be
\label{eq:ratioR}
R=\frac{\sigma_{D_1}^2}{\sigma_{D_1}^2-\sigma_{D_3}^2}
\ee
However, with our values of $D_1$ and $D_3$, the observed differences $\sigma_{D_1}^2-\sigma_{D_3}^2$ are small and sensitive to instrumental noise source. Relative errors on $\lo$ deduced by this method can attain more than 100\% in typical observing conditions, and it did not give satisfactory results on our data sets.

Here we propose another estimator to derive $\lo$, based on the ratio of absolute to differential variances of image motion:
\be
\label{eq:ratioQ}
Q_i=\frac{\sigma_{D_i}^2}{\sigma_{\mbox{\scriptsize diff}}^2}
\ee
where $\sigma_{D_i}^2$ is the absolute variance corresponding to the sub-pupil $i$, and $\sigma_{\mbox{\scriptsize diff}}^2$ the differential variance used to calculate the seeing. $\sigma_{\mbox{\scriptsize diff}}^2$ is calculated from images given by sub-pupils 1 and 2. Two values are available: $\sigma_{\mbox{\scriptsize diff}, l}^2$ and $\sigma_{\mbox{\scriptsize diff},t}^2$ for longitudinal and transverse variances. This gives two expressions for the ratios $Q_i$ (using Eq~\ref{eq:varabs} for $\sigma_{D_i}^2$ and eqs~5 and 8 of [\cite{Tokovinin02}] for $\sigma_{\mbox{\scriptsize diff}}^2$):
\be
\begin{array}{lll}
Q_{i,t} & = & \displaystyle \frac{\sigma_{D_i}^2}{\sigma_{\mbox{\scriptsize diff},t}^2}\; = \; 0.17 \frac{D_i^{-1/3}-1.525 \lo^{-1/3}}{0.364 D_1^{-1/3} -0.2905 B^{-1/3}}\\ \\

Q_{i,l} & = & \displaystyle\frac{\sigma_{D_i}^2}{\sigma_{\mbox{\scriptsize diff},l}^2}\; = \; 0.17 \frac{D_i^{-1/3}-1.525 \lo^{-1/3}}{0.364 D_1^{-1/3} -0.1904 B^{-1/3}}
\end{array}
\ee
Using absolute variances from the 3 sub-pupils, we get 6 estimations of $\lo$, from which we take the median value. Note that the absolute variance at the numerator of Eq.~\ref{eq:ratioQ} may be contaminated by telescope vibrations. Hence we use only the $x$ direction (declination axis) to compute absolute variances to reduce oscillations from the motor of the mount.

\begin{figure}
\includegraphics[width=8cm]{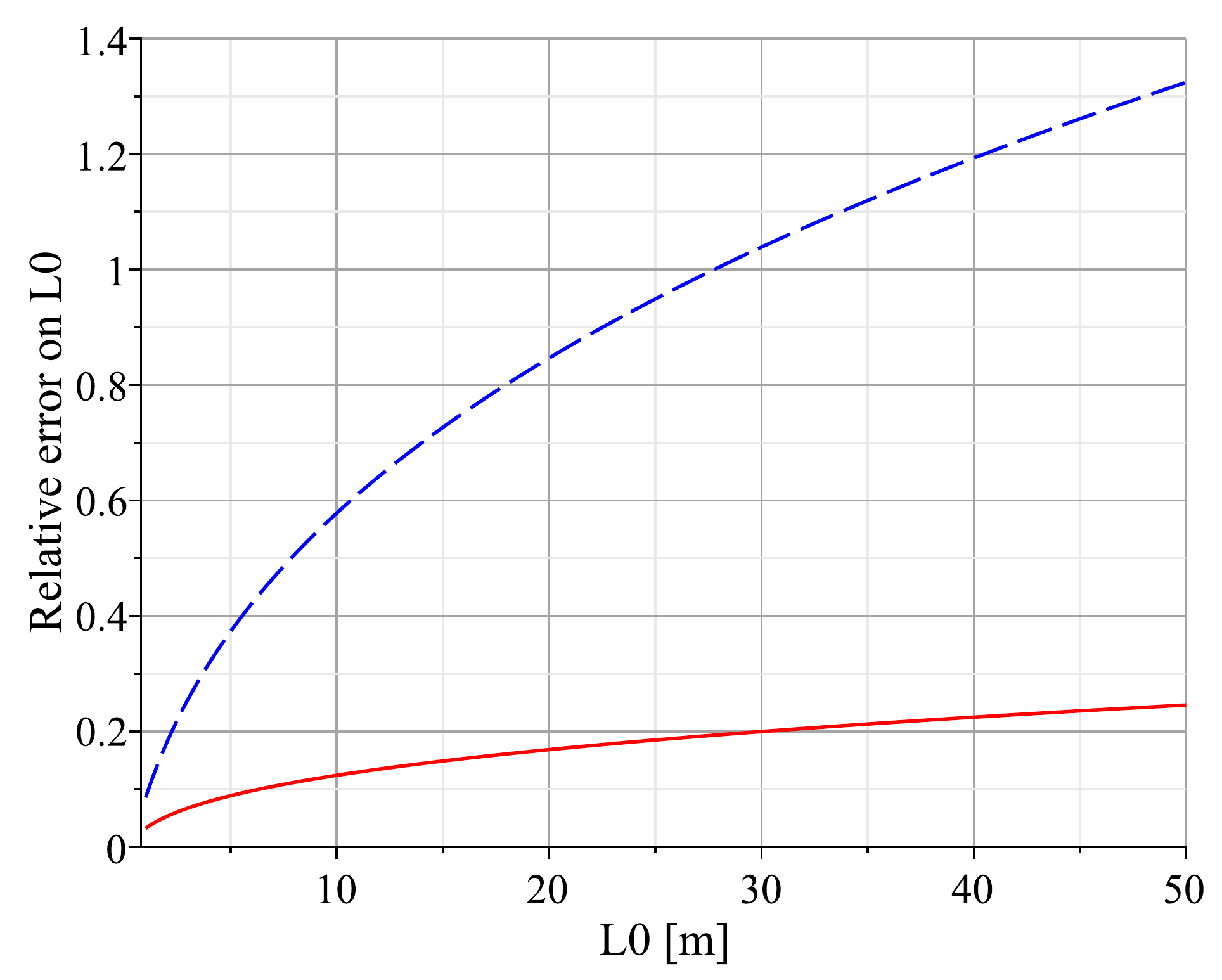}\ \includegraphics[width=9cm]{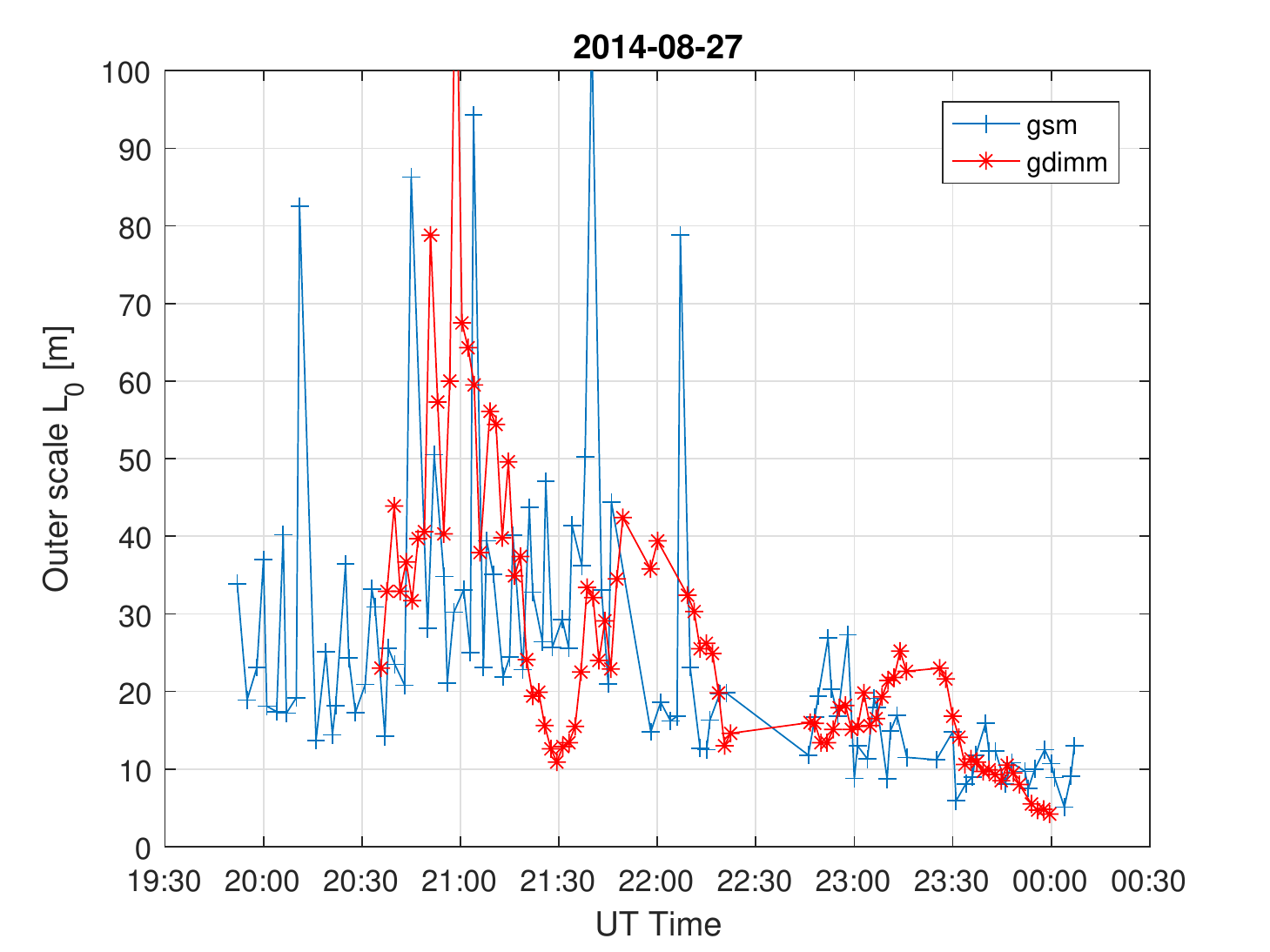} 

\caption{Left: expected relative error on $\lo$ as a function of $\lo$, assuming that the only source of uncertainty is the statistical error on variances. Dashed curve: $\lo$ deduced from the ratio $R$ (Eq.~\ref{eq:ratioR}). Solid curve: $\lo$ deduced from the ratio $Q$ (Eq.~\ref{eq:ratioQ}). These curves are calculated for an average of 8 variances (corresponding to a time interval of 16mn of data), each variance being computed from sets of $N=2000$ images. Right: Comparison of outer scale $\lo$ measured by GDIMM and GSM during the night of 2014/08/27.}\label{fig:errL0}
\end{figure}

This estimator appears to give better performances on the estimation of outer scales. And to increase  accuracy, we perform a rolling average of measured variances (they are calculated every 2mn) over time intervals of 15 to 30 minutes. Fig.\ref{fig:errL0} (left) displays the expected relative error on $\lo$, assuming that the only source of uncertainty is the statistical error on variances\cite{Sarazinroddier90}. It suggests that a relative error  $\frac{\delta \lo}{\lo}\sim 20\%$ may be obtained, combining 8 individual variances computed on data cubes of 2000 images. Of course, other error sources have to be taken in account, and this can be interpreted as a lower value of the expectable accuracy of the method.

A comparison of GDIMM and GSM outer scales was made during several observing runs in 2014 and 2015. Fig.~\ref{fig:errL0} (right) presents a time series of $\lo$ measured by the two instruments during the night of 2014/08/27. It shows a good agreement between the two datasets.

\subsubsection{Coherence time}
As for the outer scale, progress were made to estimate properly the coherence time since our first GDIMM paper\cite{Aristidi14}. In particular by increasing the framerate to $\sim$100 frames/second, and by compensating from the exposure time effect. We give here our present algorithm for the calculation of the coherence time $\tau_0$, as defined by Roddier\cite{Roddier81}
\be
\tau_0=0.31\frac{r_0}{\bar v}
\label{eq:tau0}
\ee
where $\bar v$ the effective wind speed, is a weighted average of the wind speed on the whole atmosphere. It can be estimated\cite{Ziad12, Aristidi14, Ziad17} from the temporal structure functions $D_{\alpha|\beta} (\tau)$ of the angle of arrival (AA) in the $\alpha$ (resp. $\beta$) direction (parrallel to the right ascension (resp. declination)).  This function is zero for $\tau = 0$ and saturates to a value $D_s$ for large $\tau$, and its characteristic time 
\be
\tau_{\alpha|\beta}=\frac{D_s}{e}
\ee
defines the decorrelation time of AA fluctuations in directions $\alpha$ and $\beta$. Typical values were found to be around 10~ms at Calern observatory.  
The value $D_s$ is the absolute variance of AA and may be contamined by telescope vibrations: it is indeed a source of error in the estimation of $\tau_{\alpha|\beta}$. Structure functions are compensated from finite exposure time using the same method as for the seeing:
\be
D_{\alpha|\beta} (\tau)=D_{\alpha|\beta,T} (\tau)^{n} \; D_{\alpha|\beta,2T} (\tau)^{1-n}
\ee
where $D_{\alpha|\beta,T}$ and $D_{\alpha|\beta,2T}$ are calculated on images cubes taken with exposure times of $T$ and $2T$, and $n=1.75$ (same as for seeing calculation). Extraction of $\bar{v}$ is made using eqs.~5.10, 5.11 and 5.13 of [\cite{Conan00}] taking $\kappa=e$. Some statistics for $\bar{v}$ at Calern observatory are given in Section~\ref{par:results}. The coherence time $\tau_0$ is calculated from $r_0$ and $\bar v$ using Eq.\ref{eq:tau0}.

\section{RESULTS}
\label{par:results}
\begin{figure}
\hskip -1cm \includegraphics[width=19cm]{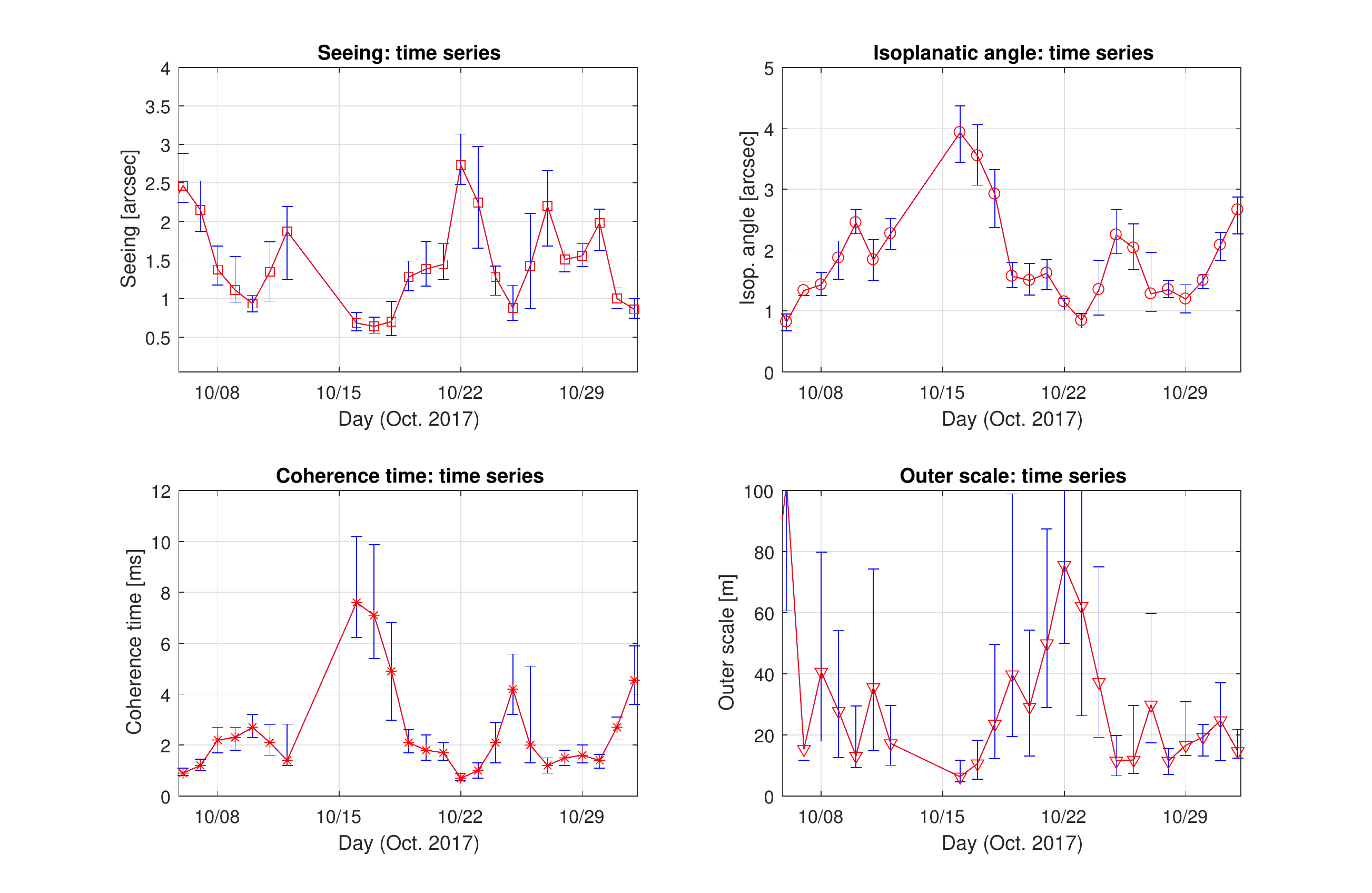} 
\caption{Time series of the 4 turbulence parameters (seeing, isoplanatic angle, coherence time and outer scale) for the month of October 2017. Each point represents the daily median value. Error bars are the interval between the first and third quartiles.}
\label{fig:params201710ts}
\end{figure}
\begin{figure}
\hskip -1cm \includegraphics[width=19cm]{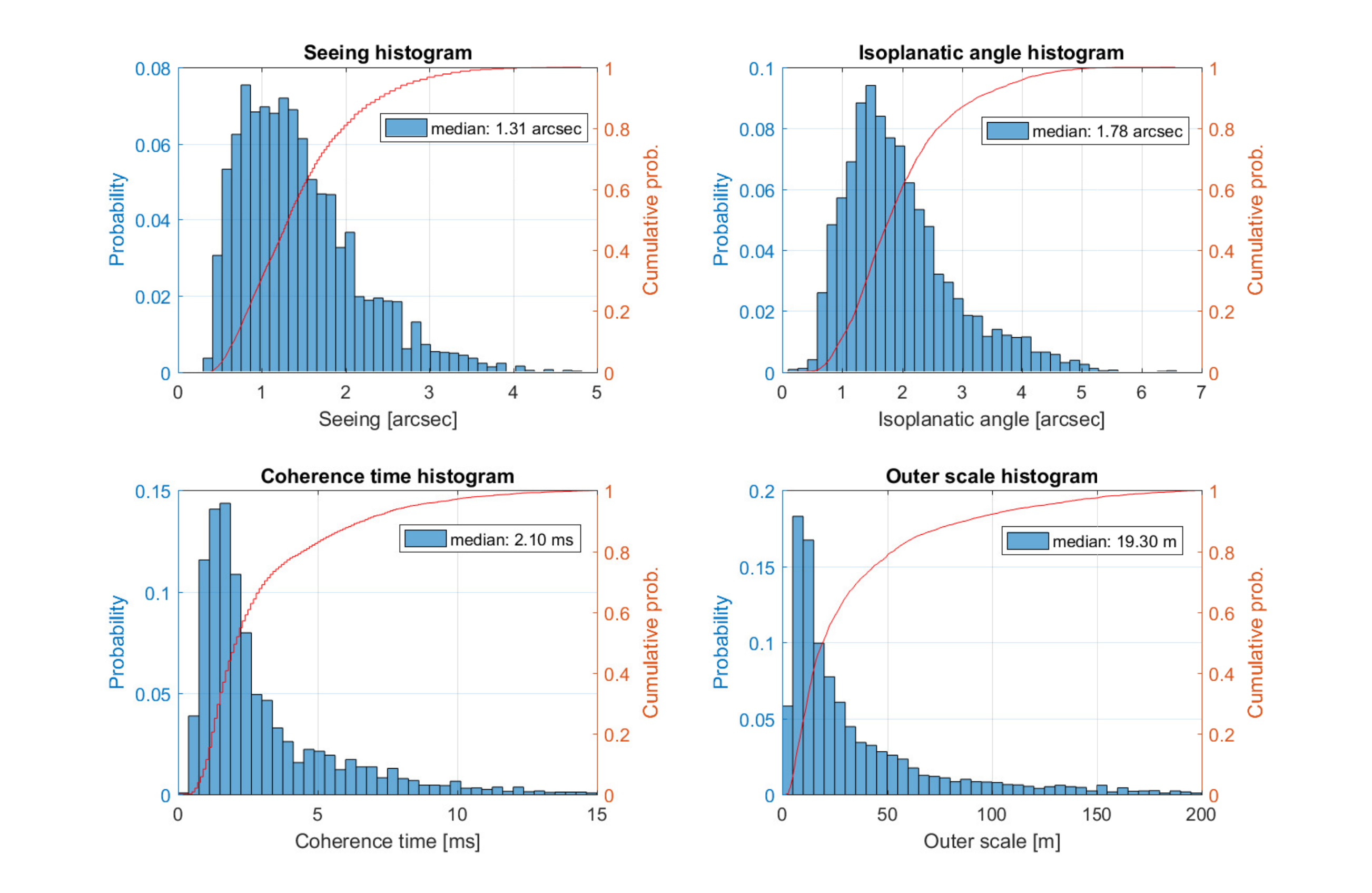} 
\caption{Histograms of turbulence parameters for the month of October 2017.}
\label{fig:params201710hist}
\end{figure}

Table \ref{table:stats} presents statistics obtained during the month of October 2017. Excellent meteo conditions were observed during this period, with 285 hours of clear night, as measured by the All-Sky camera. The total number of data points collected was 5260, representing 170~hours of data. The seeing of 1.3~arcsec is quite poor compared to other astronomical sites such as Paranal\cite{Dali09}. Indeed, and quite surprisingly, very few optical turbulence measurements were published for the site of Calern, despite its 40~years of existence. Several campaigns were made with GSM in 1999 and 2000, for a total of 15 nights. The results published in the PhD thesis of Conan~\cite{ConanThese} gave following median values: $\epsilon=0.98$~arcsec, ${\cal L}_0=27$m and $\theta_0=1.96$~arcsec, which are consistent with our measurements. Fig.~\ref{fig:params201710ts} display time series of the four main turbulence parameters ($\epsilon$, $\theta_0$, $\tau_0$ and ${\cal L}_0$). Histograms shown in Fig.~\ref{fig:params201710hist} exhibit a log-normal distribution for all quantities.

In addition to the 4 main turbulence parameters, Table \ref{table:stats} gives statistics for some additional quantities. The scintillation index is calculated from images by the pupil 3, i.e. with an aperture of 10~cm. The effective wind speed $\bar{v}\simeq 11$~m/s is about twice of the ground median wind speed $v_g=4.5$~m/s measured during the same period by the meteo station. It is lower than the value observed at Paranal, but the GDIMM data sample covers only one month and no general conclusions can be drawn from it.

\begin{figure}
\begin{center}
\hskip -1cm \includegraphics[width=17cm]{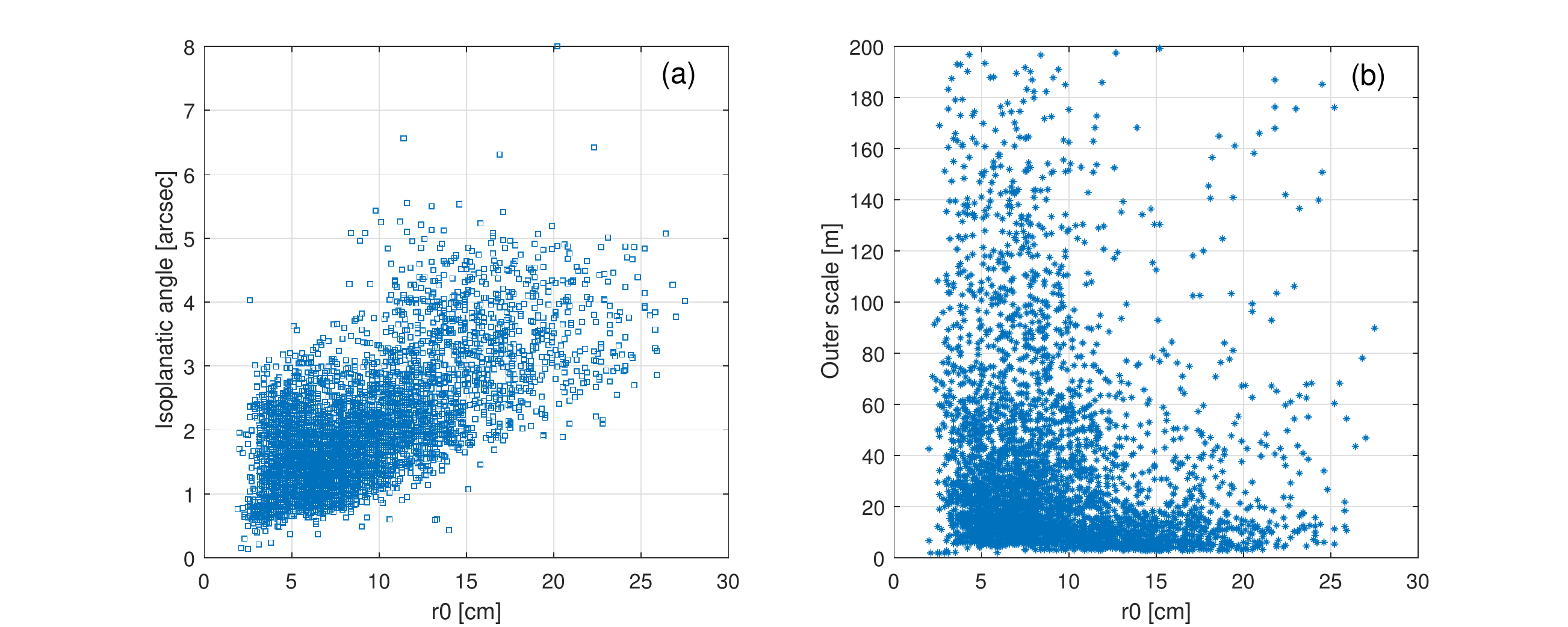} 
\end{center}
\caption{Scatter plots of isoplanatic angle (a) and outer scale (b) as a function of the Fried parameter $r_0$ for the month of October 2017.
}
\label{fig:r0l0isop}
\end{figure}

The equivalent altitude $\bar{h}~\simeq 3$~km suggests that the turbulence is spread over the whole atmosphere, high altitude layers having a significant contribution to the total integrated turbulence. This is confirmed by PML turbulence profile measurements\cite{Benrahhal18}, as well by the correlation between the Fried parameter $r_0$ and the isoplanatic angle $\theta_0$ (Fig.~\ref{fig:r0l0isop}a). Indeed $\theta_0$ involves an integral of the turbulence over the whole atmosphere wheighted by a factor $h^{5/3}$ (with $h$ the altitude, see eq.~11 of~[\cite{Looshogge79}]), and is dominated by high-altitude layers. 
\begin{table}
\begin{center}
\begin{tabular}{l|cccc||c}\hline
Parameter (Calern)                         &  Median   &   Mean &   $Q_1$  &  $Q_3$ & Paranal\cite{Dali09}\\\hline
Seeing [arcsec]                       &  1.31     &   1.44 &   0.90          &  1.80 & 0.81 \\
Isoplanatic angle [arcsec]            &  1.78     &   1.96 &   1.31             &  2.39 & 2.45 \\
Coherence time [ms]             &  2.10     &  3.00 &   1.30          & 3.7 & 2.24 \\
Outer scale [m]                    &  19.30    & 34.32 &   10.20         & 43.60 & 17.1 \\\hline
Fried parameter $r_0$ [cm]        &  7.70     &   8.84  &  5.60              & 11.20 & 12.4\\
Scintill. index [\%]         & 2.55      &  3.52 &  1.56           & 4.25 & 1.5\\
Altitude $\bar{h}$ [m]            & 2957      &  3078 &  2250           & 3818 & 3256\\
Effective wind speed [m/s]      & 11.42     &  11.23 &  8.44           & 13.9 & 17.3\\ \hline
 \end{tabular} \ \ 
\end{center}
\caption{Statistics of turbulence parameters at Calern observatory during the month of October 2017. $Q_1$ and $Q_3$ are the first and third quartile. For a sake of comparison, last column displays median values of same parameters measured at Paranal\cite{Dali09} (scintillation index, $\bar{h}$ and $\bar{v}$ were computed from median values of $\epsilon$, $\theta_0$ and $\tau_0$). }
\label{table:stats}
\end{table}

Outer scale statistics are the first published for the GDIMM instrument. They are preliminary, the algorithm to extract ${\cal L}_0$ from photocenter data is still under developpement and needs in particular more cross-calibration with GSM data. But values around 20m with a large dispersion are typical of what is found on other sites\cite{Martin02}. The scatter plot of Fig~\ref{fig:r0l0isop}b shows no correlation between ${\cal L}_0$ and $r_0$, as found in previous GSM campains\cite{Dali09}.


\section{CONCLUSION}
The GDIMM monitor is operational since the end of 2015, as a part of the Calern Atmospheric Turbulence Station. It is now a fully automatic instrument which provides continuous monitoring of turbulence parameters above the Calern observatory. Data are displayed in real time through a website ({\tt cats.oca.eu}), the idea being to provide a service available to all observers at Calern, as well as building a database to make long-term statistics of turbulence, still lacking for this site.

GDIMM is a small instrument, easy to transport to make measurements at any site in the world, and was designed to replace the GSM and provide a monitoring of the four integrated parameters of the atmospheric turbulence, i.e. seeing, isoplanatic angle, coherence time and outer scale. The 3 first quantities are of fundamental importance for adative optics (AO) correction: a large coherence time should allow to reduce the delay error, a small seeing value to easily close the loop and benefit from a rather good correction, and a large isoplanatic angle to reduce the anisoplanatic error, enlarge the sky coverage and allow very wide fields of correction (see [\cite{Carbillet17}] and references therein). Indeed several projects regarding adaptive optics systems are planned at the MéO and C2PU\cite{Bendjoya12} telescopes and they will benefit of the data given by the CATS station.

A true asset of GDIMM is the possibility to measure the outer scale. Obtain reliable values of ${\cal L}_0$ is a challenge with small instruments, and this parameter is often neglected, though it has a strong impact of high angular resolution techniques, especially for extremely large telescopes (see the discussion in [\cite{Ziad08}] and references therein). Our efforts will now focus on algorithms to compute ${\cal L}_0$ and $\tau_0$, in particular to cross-check them with other site-testing instruments, such as GSM, to obtain values as reliable as possible. A more detailed paper presenting in particular the error budget on all parameters is in preparation.

\section{Acknowledgments}
%
%
We would like thank Jean-Marie Torre and Herv\'e Viot, from the Calern technical staff, for their valuable help on the electronics of the instrument. The CATS project has been done under the financial support of CNES,
Observatoire de la C\^ote d'Azur, Labex First TF, AS-GRAM, Federation
Doblin, Universit\'e de Nice-Sophia Antipolis and R\'egion Provence Alpes
C\^ote d'Azur. 

\bibliography{biblio}   
\bibliographystyle{spiebib}   

\end{document}